\documentclass[a4paper]{article}
\usepackage[utf8]{inputenc}
\usepackage[T1]{fontenc}
\usepackage{lmodern}
\usepackage{fullpage}
\usepackage{amsmath,amssymb,array,mathrsfs,stmaryrd,amsthm}
\usepackage{booktabs}
\usepackage{enumitem}
\usepackage{hyperref}
\usepackage{graphicx}

\newtheorem{theorem}{Theorem}
\theoremstyle{plain}
\newtheorem{example}[theorem]{Example}

\title{\textbf{BeyondBenford}: An R Package to Determine Which of Benford's or BDS's Distributions is the Most Relevant}
\author{by Stéphane Blondeau Da Silva}
\date{}

\begin{document}

\maketitle

\abstract{The package \textbf{BeyondBenford} compares the goodness of fit of Benford’s and Blondeau Da Silva’s (BDS's)
digit distributions in a dataset. The package is used to check whether the data distribution is consistent with theoretical distributions highlighted 
by Blondeau Da Silva or not: this ideal theoretical distribution must be at least approximately followed by the data for the use of BDS’s model to 
be well-founded. It also allows to draw histograms of digit distributions, both observed in the dataset and given by the two theoretical approaches. 
Finally, it proposes to quantify the goodness of fit via Pearson’s chi-squared test.}

\section{Introduction}

Benford's Law, also called Newcomb-Benford's Law, is somewhat surprising; indeed, the first digit $d$, $d\in\llbracket1,9\rrbracket$, of 
numbers in many naturally occurring collections of data does not follow a discrete uniform distribution, as might be thought, but a logarithmic 
distribution (see the recent books of \cite{mi} and \cite{bh}). Discovered by the astronomer Newcomb in $1881$ \cite{new}, it was 
definitively brought to light by the physicist Benford in $1938$ \cite{ben}. The probability that $d$ is the first digit of a number is approximately: 
\begin{align*}
\log(1+\frac{1}{d})\enspace.
\end{align*} 
It can also be extended to digits beyond the first one \cite{hit}, the probability for $d$, $d\in\llbracket0,9\rrbracket$, to be the $p^{\text{th}}$ 
digit of a number being: 
\begin{align*}
\sum_{j=10^{p-2}}^{10^{p-1}-1}\log(1+\frac{1}{10j+d})\enspace.
\end{align*}

It was quickly admitted that numerous empirical data sets follow Benford's law: economic data \cite{seh}, social data \cite{gol}, demographic data \cite{NW,lee}, 
physical data \cite{knu,BK,ni,ale} or biological data \cite{cos,FGP} for instance; to such an extent that this law was used to detect possible frauds in lists of 
socio-economic data \cite{var,nig,dur,sav,to,rau} or in scientific publications \cite{alv}. 

Nevertheless many discordant voices brought a significantly different message. By putting aside the distributions known to fully disobey Benford's law 
\cite{rai,hill,tol,sco,bee,dec}, this law often appeared to be a good approximation of the reality, but no more than an approximation \cite{sco,
sav,dec,gau,goo}. 

Similar to first digit case, the distributions of digits beyond the first have been observed in various application areas \cite{gey,ale,alv} and have also been used 
to detect frauds \cite{car,tho,meb,cho,die,joe}. Once more, limits of such methods were emphasized \cite{meb,cho,die}. 

Blondeau Da Silva, considering data as realizations of a homogeneous and expanded range of random variables following discrete uniform distributions, showed that, the 
proportion of each $d$, $d\in\llbracket0,9\rrbracket$, as leading digit \cite{blo} or as other digit \cite{blon} structurally fluctuates. He demonstrated that, 
in his models, the predominance of $1$ as digit (followed by $2$ and so on) is all but surprising, and that the observed fluctuations around the values of 
probability determined by Benford's Law are also predictible: there is not a single Benford's Law but numerous distinct laws each of them determined by a parameter, 
the upper-bound of the considered data.

The huge and growing literature on Benford’s law is available on the online database \url{www.benfordonline.net} with well over $1000$ papers. Two CRAN Packages
already enable to check whether datasets conform to Benford's law or not: \textbf{BenfordTests} and \textbf{benford.analysis}. The package \textbf{BeyondBenford} 
compares the goodness of fit of Benford’s Law, on the one hand, and BDS’s Laws, on the other hand. Indeed, these latter, under certain conditions that we will recall, 
allow a better reliability of adjustment. The package \textbf{BeyondBenford} calculates the digit distribution in the considered dataset and determines whether it is 
consistent with BDS's or Benford's one. It also provides plotting tools for the visual evaluation of these distributions. We will walk through a detailed
example to give an overview of the \textbf{BeyondBenford} package.

\section{An example to get familiar with the main functions}

Street addresses of Pierre-Buffiere, a small town of approximately $1200$ inhabitants in Haute-Vienne (France), are available on:
\begin{center}
\url{www.data.gouv.fr/fr/datasets/base-d-adresses-nationale-ouverte-bano/}\,,
\end{center}
which is an open platform for French public data. 

After loading the package ({\fontfamily{lmtt}\selectfont library(BeyondBenford)}), the code {\fontfamily{lmtt}\selectfont data(address\textunderscore PierreBuffiere)} 
gives us access to the sample data. This factor contains $346$ rows, with each row representing an address number.

\section{Are the data consistent with BDS's model?}

This is the first essential question that must be answered in the affirmative. If this is not the case, comparisons are not relevant: the package should not be used.
Indeed the use of the package is appropriate when the studied data can be considered as realizations of a homogeneous and expanded range of random 
variables approximately following discrete uniform distributions. In this model, the data is strictly positive and is upper-bounded, 
constraint which is often valid in datasets, the physical, biological, demographic, social and economical quantities being limited \cite{blo}. 

Among the different domains studied by Benford \cite{ben}, some could be well adapted to our model: sizes of populations or street addresses for example 
(see \cite{blo} for a detailed explanation). \cite{jan} advised precisely to use their own similar model in the case of street addresses or when considering the 
first-page numbers of articles in a bibliography. 

\cite{blon} showed that the model induces a specific distribution of positive integers determined by an upper-bound. Hence, in order to conform as closely as possible 
to the model, the studied database must have a distribution similar to that described in \cite{blon}. In $p^{\text{th}}$ digit case, the probability $p_k$ to obtain 
the number $k\in\llbracket10^{p-1};u_b\rrbracket$ (where $u_b$ is the upper-bound) verifies:
\begin{align*}
p_k=\frac{1}{u_b-10^{p-1}+1}\sum_{i=k+1-10^{p-1}}^{u_b+1-10^{p-1}}\frac{1}{i}\enspace.
\end{align*}

\begin{figure}[ht]
\centering
\includegraphics[scale=0.75,clip=true]{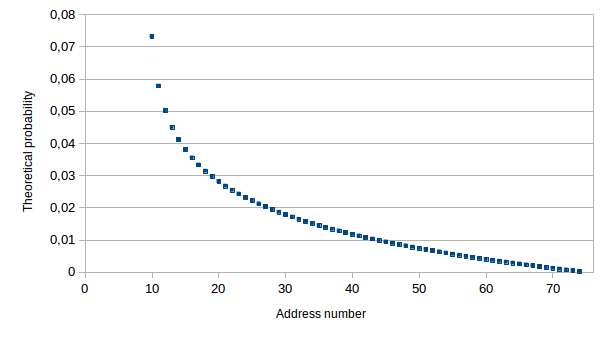}
\caption{Theoretical distribution of street numbers in the second digit case, the upper-bound being $74$.}\label{fig0}
\end{figure}

In the studied example, the maximum value of the street number is $74$. The associated theoretical distribution for the second digit is plotted in Figure \ref{fig0}.

The \textbf{BeyondBenford} package provides plotting tools to determine whether the data is consistent with BDS's model: the function {\fontfamily{lmtt}\selectfont dat.distr}.  The function’s 
arguments are as follows:
\begin{enumerate}[label=$\circ$]
\item \textbf{dat}: the considered dataset, a data frame containing non-zero real numbers.
\item \textbf{xlab}: the $x$-axis label. Default value: {\fontfamily{lmtt}\selectfont xlab="data"}.
\item \textbf{ylab}: the $y$-axis label. Default value: {\fontfamily{lmtt}\selectfont xlab="Frequency"}.
\item \textbf{main}: the title of the graph. Default value: {\fontfamily{lmtt}\selectfont main="Distribution of data"}.
\item \textbf{theor}: if {\fontfamily{lmtt}\selectfont theor=TRUE} BDS’s theoretical distribution is plotted, otherwise
only the histogram is represented. Default value: {\fontfamily{lmtt}\selectfont theor=TRUE}.
\item \textbf{nclass}: a strictly positive integer: the number of classes in the histogram.
Default value: {\fontfamily{lmtt}\selectfont nclass=50}.
\item \textbf{col}: the color used to fill the bars of the histogram. {\fontfamily{lmtt}\selectfont NULL} yields unfilled bars.
Default value: {\fontfamily{lmtt}\selectfont col="lightblue"}.
\item \textbf{conv}: if {\fontfamily{lmtt}\selectfont conv=1}, all values of the dataset are multiplied by $10^k$ where $k$ is
the smallest positive integer such that all non-zero numerical values in the newly multiplied 
data frame have an absolute value greater than or equal to $1$. Default value: {\fontfamily{lmtt}\selectfont conv=0}.
\item \textbf{upbound}: a positive integer, which characterizes the data. All (or most) of the 
data are lower than this "upper-bound". Default value: {\fontfamily{lmtt}\selectfont upbound=ceiling(max(dat))}.
\item \textbf{dig}: the chosen position of the digit (from the left). Default value: 
{\fontfamily{lmtt}\selectfont dig=1}.
\item \textbf{colt}: the color used to plot BDS’s theoretical distribution. 
Default value: {\fontfamily{lmtt}\selectfont colt="red"}.
\item \textbf{ylim}: a two-components vector: the range of $y$ values. Default value: 
{\fontfamily{lmtt}\selectfont ylim=NULL}.
\item \textbf{border}: the color of the border around the bars. Default value: 
{\fontfamily{lmtt}\selectfont border="blue"}.
\item \textbf{nchi}: the number of classes for values from $10^{p-1}$ to 
max(max(data),upbound). If {\fontfamily{lmtt}\selectfont nchi>0}, the function returns the chi-squared statistic 
(with $nchi-1$ degrees of freedom) of goodness of fit determined by the different classes. The 
null hypothesis states that the studied distribution is consistent with the considered 
theoretical distribution. Default value: {\fontfamily{lmtt}\selectfont nchi=0}.
\item \textbf{legend}: if {\fontfamily{lmtt}\selectfont legend=TRUE}, the legend is displayed.
Default value: {\fontfamily{lmtt}\selectfont legend=TRUE}.
\item \textbf{bg.leg}: the background color for the legend box. Default value: 
{\fontfamily{lmtt}\selectfont bg.leg="gray85"}.
\end{enumerate}

\begin{figure}[ht]
\centering
\includegraphics[scale=0.5,clip=true]{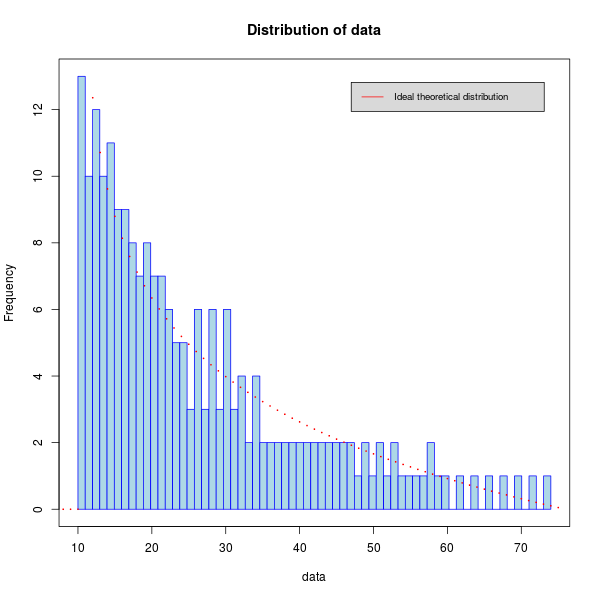}
\caption{The data histogram of street numbers in the second digit case and its BDS’s theoretical distribution.}\label{fig1}
\end{figure}

Let us apply the {\fontfamily{lmtt}\selectfont dat.distr} function to the {\fontfamily{lmtt}\selectfont address\textunderscore PierreBuffiere} dataset. The output from {\fontfamily{lmtt}\selectfont dat.distr} is 
the data histogram along with optional BDS’s theoretical distributions (Figure \ref{fig1}).

\begin{example}
{\fontfamily{lmtt}\selectfont \#\#} Both the histogram and theoretical distribution are represented\\
{\fontfamily{lmtt}\selectfont dat.distr(address\_PierreBuffiere, dig=2, nclass=65)}
\end{example}

Note that, out of the $346$ values, only $217$ are taken into consideration here because the values need to have at least $2$ digits. 

The data distribution looks similar to the one described by BDS's model (Figure \ref{fig1}). Let us provide a second example in which we numerically determine whether 
the studied distribution is conform to the theoretical distribution or not:
\begin{example}
{\fontfamily{lmtt}\selectfont 
\#\#} The function returns the chi-squared statistic of goodness of fit 
determined by nchi classes.\\
{\fontfamily{lmtt}\selectfont dat.distr(address\_PierreBuffiere, dig=2, nchi=4)\\
$[$1$]$ "Class freq.:" "130"       "51"       "25"	\\
$[$5$]$ "11"\\
$[$1$]$ "Theor. freq.:"    "127.050977504473"     "54.6446927640259"     "26.7009825219254"	\\
$[$5$]$ "8.60334720957567"\\
\hspace*{2.32cm}chi2     \hspace*{2.06cm}        pval \\
1   \hspace*{0.36cm}Chi2 value is:   \hspace*{0.21cm}The p-value is:\\
2 1.08754607580421 0.780081329402347}
\end{example}

The {\fontfamily{lmtt}\selectfont dat.distr} function returns, if requested, the frequencies of each equal-sized class of the dataset and the associated theoretical frequencies. It also returns 
a data frame containing the value of the chi squared test-statistic and its p-value. Note that the number of classes is limited by the theoretical frequencies that 
cannot exceed $5$ in Pearson's chi-squared test \cite{pea}. In our example, the null hypothesis cannot be rejected: the studied distribution is consistent with the 
theoretical distribution.

\section{Comparisons of the goodness of fit of Benford’s and BDS's digit distributions in a dataset}

\subsection{Raw data}

The package \textbf{BeyondBenford} provides a function returning the frequencies of each figure at a given position in the considered dataset: 
{\fontfamily{lmtt}\selectfont obs.numb.dig}. Its two arguments \textbf{dat} and \textbf{dig} have already been defined in {\fontfamily{lmtt}\selectfont dat.distr} function. The function output is a 
vector containing the frequencies of each figure in ascending order.

Let us give an example:
\begin{example}
{\fontfamily{lmtt}\selectfont obs.numb.dig(address\_PierreBuffiere, dig=2)\\
$[$1$]$ 31 24 27 21 25 17 21 16 19 16}
\end{example}

For instance, there are $25$ values with the second digit being a $4$.

The package \textbf{BeyondBenford} also provides a function returning Benford’s probability that a figure is at a given position: {\fontfamily{lmtt}\selectfont Benf.val}. 
Its argument \textbf{dig} has already been defined in {\fontfamily{lmtt}\selectfont dat.distr} function. Its second argument \textbf{fig} stands for the considered figure.

Let us give an example:
\begin{example}
{\fontfamily{lmtt}\selectfont Benf.val(4, dig=2)\\
$[$1$]$ 0.1003082}
\end{example}

The package \textbf{BeyondBenford} at last provides a function returning BDS’s probability that a figure is at a given position (once the
associated upper-bound has been specified): {\fontfamily{lmtt}\selectfont Blon.val}. Its three arguments \textbf{dig}, \textbf{upbound} and \textbf{fig} have already been 
defined above.

Let us give an example:
\begin{example}
{\fontfamily{lmtt}\selectfont Blon.val(fig=4, dig=2,upbound=74)\\
$[$1$]$ 0.09836642}
\end{example}

In Table \ref{tab1} below, the frequencies of each figure in the considered dataset, regarding the second digit of address numbers, are listed.
\begin{table}[ht]
\centering
\begin{tabular}{|c||c|c|c|}
\hline
Figure & Frequency in the database & BDS's values & Benford's values\\\hline
$0$&$0.1429$ & $0.1436$& $0.1197$  \\\hline
$1$&$0.1106$ & $0.1247$& $0.1139$  \\\hline
$2$&$0.1244$ & $0.1136$& $0.1088$  \\\hline
$3$& $0.0968$&$0.1053$ & $0.1043$  \\\hline
$4$&$0.1152$ &$0.0984$ & $0.1003$  \\\hline
$5$&$0.0783$ &$0.0924$ & $0.0967$  \\\hline
$6$&$0.0968$ &$0.0872$& $0.0934$  \\\hline
$7$& $0.0737$&$0.0825$ & $0.0904$  \\\hline
$8$& $0.0876$& $0.0782$& $0.0876$  \\\hline
$9$&$0.0737$ &$0.0742$ & $0.0850$  \\\hline
\end{tabular}
\caption{Values of frequency of each figure as second digit in the database, Benford's and BDS's theoretical values (the chosen upper-bound being $74$).
These values are rounded to the nearest ten-thousandth.}\label{tab1}
\end{table}

The BDS's theoretical values seem slightly better; in particular the frequency range is higher, both for the observed data and for BDS's theoretical values. 

\subsection{Plotting tools}

The \textbf{BeyondBenford} package provides plotting tools to perform the comparison between the two models with the function {\fontfamily{lmtt}\selectfont digit.distr}. In addition to
arguments that are shared with {\fontfamily{lmtt}\selectfont dat.distr} (\textbf{dat}, \textbf{dig}, \textbf{upbound}, \textbf{main}), the {\fontfamily{lmtt}\selectfont digit.distr} function has the following 
additional arguments:
\begin{enumerate}[label=$\circ$]
\item \textbf{mod}: if {\fontfamily{lmtt}\selectfont mod="ben"}, the data histogram and that of Benford are displayed, 
if {\fontfamily{lmtt}\selectfont mod="ben$\&$blo"}, the data histogram, that of Benford and that of BDS are 
plotted, and otherwise the data histogram and that of BDS are given. 
Default value: {\fontfamily{lmtt}\selectfont mod="ben"}.
\item \textbf{col}: a vector containing two colors used to fill the bars of the histogram, 
if {\fontfamily{lmtt}\selectfont mod="ben"}. Default value: {\fontfamily{lmtt}\selectfont col=c("\#FFFFAA", "\#AAFFAA")}.
\item \textbf{colbebl}: a vector containing three colors used to fill the bars of the histogram, 
if {\fontfamily{lmtt}\selectfont mod="ben$\&$blo"}. Default value: {\fontfamily{lmtt}\selectfont colbebl=c("\#FFFFAA", "\#AAFFAA", "\#AAFFFF")}.
\item \textbf{colbl}: a vector containing two colors used to fill the bars of the histogram, 
if the latter case. Default value: {\fontfamily{lmtt}\selectfont colbl=c("\#FFFFAA", "\#AAFFFF")}.
\item \textbf{legend}: if {\fontfamily{lmtt}\selectfont legend=TRUE}, the legend is displayed. Default value: 
{\fontfamily{lmtt}\selectfont legend=TRUE}.
\item \textbf{leg}: a two-components vector containing text appearing in the legend, if 
{\fontfamily{lmtt}\selectfont mod="ben"}. Default value: {\fontfamily{lmtt}\selectfont leg=c("Observed", "Benford")}.
\item \textbf{legbebl}: a three-components vector containing text appearing in the legend, 
if {\fontfamily{lmtt}\selectfont mod="ben$\&$blo"}. Default value: {\fontfamily{lmtt}\selectfont legbebl=c("Observed", "Benford", "Blondeau")}.
\item \textbf{legbl}: a two-components vector containing text appearing in the legend, 
if the latter case. Default value: {\fontfamily{lmtt}\selectfont legbl=c("Observed", "Blondeau")}.
\end{enumerate}

\begin{figure}[ht]
\centering
\includegraphics[scale=0.5,clip=true]{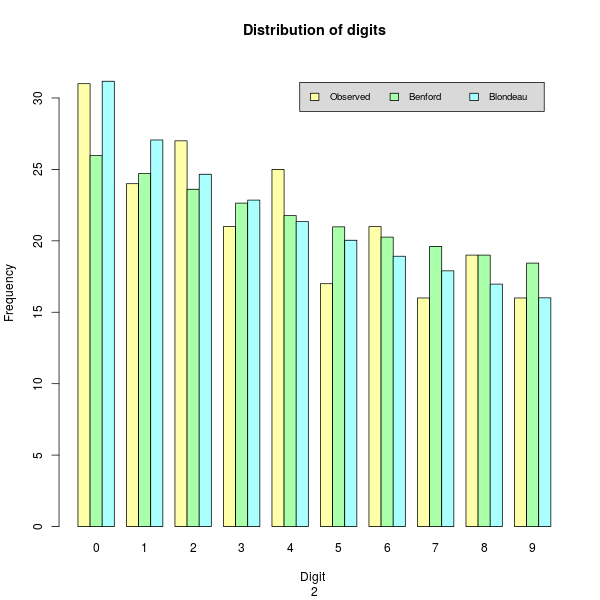}
\caption{The histogram of Pierre Buffière's street numbers in the second digit case along with Benford's and BDS’s distributions of digit.}\label{fig2}
\end{figure}

Let us apply the {\fontfamily{lmtt}\selectfont digit.distr} function to the {\fontfamily{lmtt}\selectfont address\textunderscore PierreBuffiere} dataset. The output from {\fontfamily{lmtt}\selectfont dat.distr} is 
a histogram of theoretical and experimental digit distribution (Figure \ref{fig2}). 

\begin{example}
{\fontfamily{lmtt}\selectfont digit.distr(address\_PierreBuffiere, dig=2, mod="ben$\&$blo")}
\end{example}

Naturally, Figure \ref{fig2} is consistent with Table \ref{tab1}.

\subsection{Pearson's chi-squared test}

To quantify the quality of theorical models, we use Pearson's chi-squared test of goodness of fit \cite{pea}: the 
null hypothesis states that the studied distribution is consistent with the considered theoretical distribution, \textit{i.e.} Benford's or BDS's ones. The function 
{\fontfamily{lmtt}\selectfont chi2} determines the test statistic and its associated p-value. In addition to arguments that are shared with {\fontfamily{lmtt}\selectfont dat.distr} (\textbf{dat}, \textbf{dig}, 
\textbf{upbound}), the {\fontfamily{lmtt}\selectfont chi2} function has the following specific arguments:
\begin{enumerate}[label=$\circ$]
\item \textbf{mod}: if {\fontfamily{lmtt}mod="ben"}, the theorical distribution considered is that of Benford, else it is
BDS’s ones which is chosen. Default value: {\fontfamily{lmtt}\selectfont mod="ben"}.
\item \textbf{pval}: if {\fontfamily{lmtt}pval=0}, the p-value is not returned, else it is available. Default value: {\fontfamily{lmtt}\selectfont pval=0}.
\end{enumerate}

Let us apply this function to the {\fontfamily{lmtt}\selectfont address\textunderscore PierreBuffiere} dataset. The output from {\fontfamily{lmtt}\selectfont chi2} is 
a data frame containing the Pearson's chi-squared statistic (and the associated p-value if requested). 

\begin{example}
{\fontfamily{lmtt}\selectfont \#\#} Measure of Benford's Law goodness of fit\\
{\fontfamily{lmtt}\selectfont 
chi2(address\_PierreBuffiere, dig=2, pval=1)\\
\hspace*{2.32cm}chi2     \hspace*{2.06cm}        pval \\
1   \hspace*{0.36cm}Chi2 value is:   \hspace*{0.21cm}The p-value is:\\
2 3.84793221030181 0.921132993758269}\vspace{0.5\baselineskip}

{\fontfamily{lmtt}\selectfont \#\#} Measure of BDS's Law goodness of fit\\
{\fontfamily{lmtt}\selectfont 
chi2(address\_PierreBuffiere, dig=2, pval=1, mod="BDS")\\
\hspace*{2.32cm}chi2     \hspace*{2.06cm}        pval \\
1   \hspace*{0.36cm}Chi2 value is:   \hspace*{0.21cm}The p-value is:\\
2 2.47996278328848 0.981417506807756}
\end{example}

In both cases the null hypothesis cannot be rejected: the studied distribution is consistent with the theoretical distributions. It can be noted that 
the quality of the adjustment seems slightly better with BDS's model.

\section{Conclusion}

The use of Benford's Law has increased rapidly in the last few years in extremely diverse fields such as mathematics, physics, biology, economics and demography, to name 
but a few. But the adjustment proposed by Benford is often only approximate. In some precisely described cases, it is BDS's probability distributions that are 
preferred. Indeed the probabilies of occurrence of digits in these distributions fluctuate around Benford's values \cite{blo,blon}.

The \textbf{BeyondBenford} package is thus a relevant tool to compare the goodness of fit of Benford's and BDS's distributions in a given collection of data and to  
offer new laws to find a better approximation of digits distribution in the considered dataset.

\bibliography{blondeau-da-silva-stephane}

\begin{thebibliography}{CLRTFM08}
\bibitem[Pea00]{pea}
K.~Pearson.
\newblock On the criterion that a given system of deviations from the probable
  in the case of a correlated system of variables is such that it can be
  reasonably supposed to have arisen from random sampling.
\newblock {\em Philosophical Magazine}, 50(302):157--175, 1900.
\end{thebibliography}

\end{document}